\documentclass{ws-p8-50x6-00}

\def\lapproxeq{\lower .7ex\hbox{$\;\stackrel{\textstyle
<}{\sim}\;$}}
\def\gapproxeq{\lower .7ex\hbox{$\;\stackrel{\textstyle
>}{\sim}\;$}}
\def\newblock{ }

\begin{document}


\title{Soft interactions}

\author{Ralph Engel}

\address{Bartol Research Institute, University of Delaware\\
217 Sharp Lab, Newark DE 19716, USA\\ 
E-mail: eng@lepton.bartol.udel.edu}

\maketitle

\abstracts{
A brief introduction to the theory and phenomenology of
soft interactions is given, focusing on total and elastic 
cross sections and multiparticle production.
}


\section{Introduction}

The description of soft hadronic multiparticle production within QCD 
is one of the long-standing, unsolved problems.
In recent years progress has been made 
in understanding soft QCD processes in lattice
calculations\cite{Sachrajda01a} and also the
modified leading-log approximation proved to be a remarkable powerful
tool\cite{Khoze01a}. 
%
%
In this article we focus on another method which
is the investigation of predictions implied by
unitarity and analyticity of scattering amplitudes and general features of
gauge field theories. 
We will discuss basic concepts and predictions of general nature 
rather than one or several models in detail, emphasizing a pedagogical
presentation of the underlying  ideas. The
aim is to supply the reader with a basic theoretical framework to
understand the more specific contributions to this conference. More detailed
discussions of soft interactions and related topics can be 
found in other reviews
\cite{Kaidalov79,Collins82a,Capella94a,Abramowicz97b,Levin98b,Kaidalov99}.

\section{Regge amplitude}

\begin{figure}[htb]
\centerline{\epsfig{figure=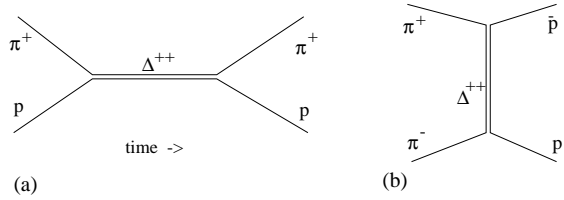,height=2.5cm}}
\caption{Production and decay of $\Delta^{++}$ in the
$s$-channel (a), and a interaction obtained by crossing 
with $\Delta^{++}$ in the
$t$-channel (b).\label{fig-res}
}
\end{figure}

Cross sections and amplitudes at low energy are fairly well understood
in terms of production and decay of a single or a few resonances, as
shown in Fig.~\ref{fig-res}a. With
increasing collision energy more and more resonances (with increasing
mass and decay width) contribute to the total cross section. This seems
to render any approach, which is based on summing all 
resonance contributions to extrapolate to high energy, infeasible. 
However, under
certain conditions this summation can be carried out and the Reggeon
amplitude is obtained \cite{Collins77}. 

Because of angular momentum conservation it is convenient to expand the 
scattering amplitude $A(s,t)$ in terms of partial waves
$a_l(s)$ for fixed angular momentum $l$
\begin{equation}
A(s,t) = 16 \pi \sum_{l=0}^{\infty} (2l+1) a_l(s) P_l(\cos \theta),
\label{partial-sum}
\end{equation}
with $\theta$ being the scattering angle in c.m.\ frame and $P_l$ the
Legendre functions of the first kind\footnote{For
sake of clarity we omit here all spin-related complications.}. For resonance
exchange the amplitude is of the well known Breit-Wigner form
$a_l \sim 1/(s-m_l^2+im_l\Gamma_l)$. 

One of the important assumptions in Regge theory is that of
maximum analyticity of the scattering amplitude, which holds in
gauge field theories. 
The amplitude is an analytic function and has only
physics-implied singularities. Thus, if the amplitude is
completely known for a given scattering process, then the amplitudes for
all other channels, which are related to the former one by crossing, 
can be obtained by analytic continuation of this amplitude. An example
for this is given in Fig.~\ref{fig-res}b. After crossing from the
$s$-channel to the $t$-channel
Eq.~(\ref{partial-sum}) reads
\begin{equation}
A(s,t)=16\pi\sum_l (2l+1) a_l(t) P_l(z_t),\hspace*{1cm} z_t = \cos
\theta_t = \frac{2s}{t-s_0}+1\ .
\label{crossed-sum}
\end{equation}
The scale $s_0$ is related to the masses of the particles ($s_0\sim
1$GeV$^2$). The partial wave amplitude describes now the
exchange of a particle with angular momentum $l$ in the $t$-channel
$a_l(t) \sim 1/(t-m_l^2+im_l\Gamma_l)$.

\begin{figure}[htb]
\centerline{
\epsfig{figure=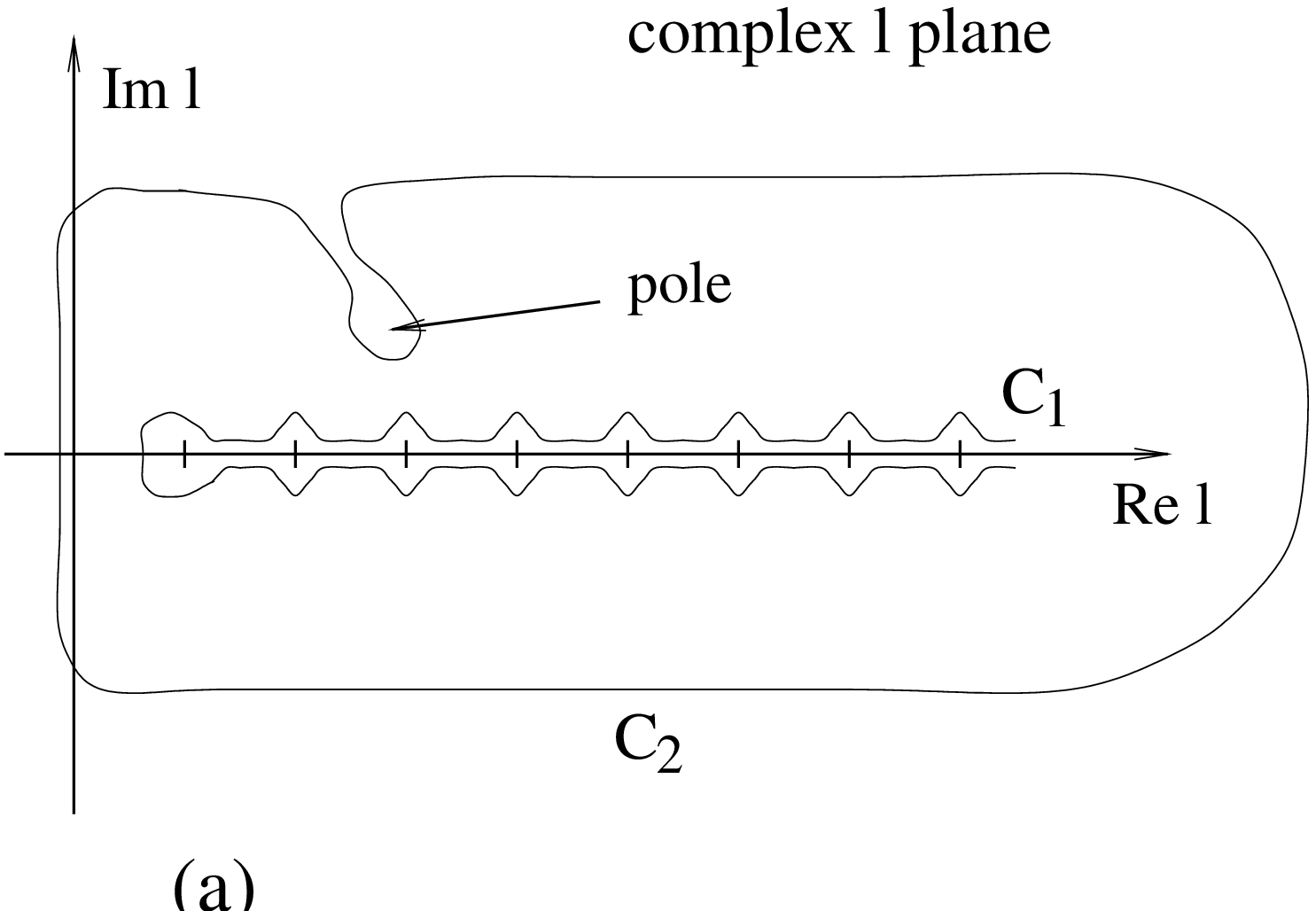,width=5.5cm}
\epsfig{figure=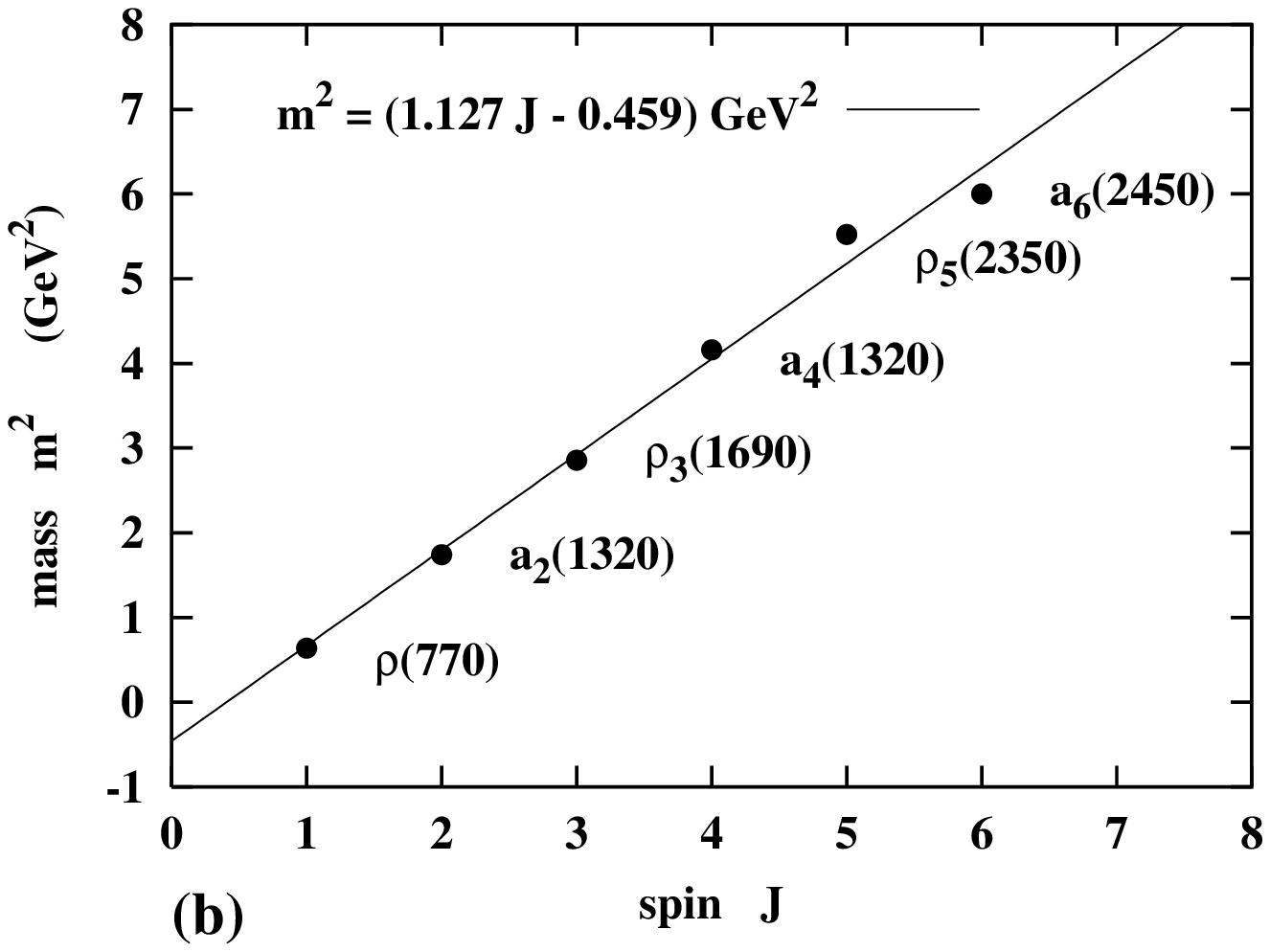,width=6.5cm}
}
\caption{(a) Integration contours for Sommerfeld-Watson
transform, and (b) mass-spin relation for
for $\rho$ and $s$ mesons (Chew-Frautschi plot). 
\label{fig-chew-frautschi}
}
\end{figure}

Using Cauchy's theorem the sum (\ref{crossed-sum}) is re-written as integral 
(Sommerfeld-Watson transformation)
\begin{equation}
A(s,t) = \sum_{\tau=\pm 1}\frac{16 \pi}{2 i} \int_{C_1} dl\; (2 l +1)
\left(\frac{1 + \tau e^{-i\pi l}}{\sin(\pi l)}\right)
\;a_l(t)\;P_l(-z_t),\hspace*{0.5cm}\tau=\pm1\ ,
\label{sommerfeld-watson}
\end{equation}
where the signature $\tau=\pm1$ separates the sum into even and odd
integer angular momenta.  The splitting
of the partial sum into positive and negative signature contributions is
needed to ensure convergence of the integral.
The integration contour runs along both sides
of the positive $\Re e\{l\}$ axis in the complex $l$ plane (see
Fig.~\ref{fig-chew-frautschi}a).

In (\ref{sommerfeld-watson}) the partial wave amplitude has to be
extended to continuous values of $l$. 
This is done by employing the experimentally
established, approximate relation between resonance mass 
and total angular momentum
$m_l^2 = a l + m^2_0$ with $a$ and $m_0$ being constants specific to the
group of resonances. Fig.~\ref{fig-chew-frautschi} shows an example for
meson resonances. From the structure of the partial wave amplitude 
\begin{equation}
a_l \sim \frac{1}{t^2 - m_l^2} =  \frac{1}{t-m_0^2-a l} 
\sim \frac{1}{l-t/a + m_0^2/a} = \frac{1}{l-\alpha(t)}
\end{equation}
follows that there is a singularity for $l=\alpha(t)$ with $\alpha(t) =
(t-m_0^2)/a$. The function $\alpha(t)$, called Regge trajectory,
is in general a nonlinear complex function of $t$ and related to the quantum
numbers of the hadrons whose mass-angular momentum relation it describes.

Assuming that $l=\alpha(t)$ is the only singularity due to $a_l(t)$ we
can displace the integration contour in Fig.~\ref{fig-chew-frautschi}a
from $C_1$ to $C_2$. The behaviour of the integrand leads to a vanishing
contribution from the semi-circle at infinity. Neglecting the
contribution from $\Im m\{l\}=-1/2$ all that is left is the contribution of
the pole at $l=\alpha(t)$. Thus the final amplitude can be written as
\begin{equation}
A(s,t) = - \frac{1+\tau e^{-i\pi\alpha(t)}}{\sin(\pi\alpha(t))}
\beta(t) P_{\alpha(t)}(-z_t)\ ,
\end{equation}
with $\beta(t)$ standing for the residue and other unimportant factors.

Finally, the Regge amplitude is obtained by going to the high-energy
limit $s \gg s_0$
\begin{equation}
P_{\alpha_k(t)}\left(-\frac{2 s}{t - s_0}-1\right) \stackrel{s
\rightarrow \infty}{\longrightarrow}
\left(\frac{s}{s_0}\right)^{\alpha_k(t)}
\end{equation}
and reads
\begin{equation}
A(s,t) = \sum_k \eta(\alpha_k(t))
\beta_k(t) \left(\frac{s}{s_0}\right)^{\alpha_k(t)}\ .
\label{regge-amp}
\end{equation}
The index $k$ represents the sum over all contributing Regge trajectories
and $\eta$ is the signature factor of the Regge trajectory $k$
\begin{equation}
\eta(\alpha_k(t)) = -\frac{1+\tau
e^{-i\pi\alpha_k(t)}}{\sin(\pi\alpha_k(t))}\ .
\end{equation}
A single term in the sum (\ref{regge-amp}) describes the contribution of an
infinite number of particles exchanged in the $t$-channel.
Each of these groups of particles is represented in the
amplitude by a quasi-particle, called reggeon, with the same quantum 
numbers.


\section{Cross section phenomenology}


\subsection{Regge-based fits}

Knowing the Regge trajectories from the measured masses of mesons and
baryons, Eq.~(\ref{regge-amp}) can be
used to predict the energy-dependence of elastic and total cross sections
\begin{equation}
\frac{d\sigma_{\rm ela}}{dt} = \frac{1}{16 \pi s^2} \left|
A(s,t)\right|^2\ ,  \hspace*{1cm}
\sigma_{\rm tot} = \frac{1}{s}\ \Im m \left( A(s, t=0)\right).
\label{cross-sections}
\end{equation}
The last relation employs the optical theorem and follows from
unitarity\footnote{The masses of the scattering particles are neglected.}. 
Using furthermore the parametrization 
$\beta^2(t)= g^2 \exp\{B_0 t\}$ and
considering only the highest Regge trajectory, one gets with
$\Delta=\alpha(0)-1$
\begin{eqnarray}
\sigma_{\rm ela} &=& (1+\rho^2) \frac{g^2 }{16 \pi}
\left(\frac{s}{s_0}\right)^{2\Delta} \exp\left\{ B_{\rm ela} t \right\},
\hspace*{1cm} \rho = \frac{\Re e\;A(s,t)}{\Im m\;A(s,t)}\bigg|_{t=0}
\nonumber\\
\sigma_{\rm tot} &=& g \left(\frac{s}{s_0}\right)^\Delta\ , 
\hspace*{1cm}
B_{\rm ela} = B_0 + 2 \alpha^\prime(0) \ln s
\label{cs-pole}
\end{eqnarray}

There are several shortcomings of the amplitude (\ref{regge-amp}). 
The total cross
section is predicted to have the energy dependence $\sigma_{\rm tot}
\sim s^\Delta$. The intercept of the highest
known Regge trajectory is about $\alpha(0)\approx 0.5$. The
experimentally observed rise of the cross sections at high energy has 
lead Pomeranchuk to propose the existence of a pole with
$\alpha(0)\gapproxeq 1$ which corresponds to the exchange of objects 
with vacuum quantum numbers. The corresponding quasi-particle is called 
the pomeron. Its trajectory
should be related to glueballs instead of mesons or baryons. So far the 
existence of glueballs could not be confirmed experimentally.

Another problem of Eq.~(\ref{regge-amp}) is the violation of unitarity
at high energy. This is most obvious by noting that the elastic cross
section is predicted to finally exceed the total one or that the
Froissart-Martin
bound $\sigma^{\rm tot} < c \ln^2s$ is violated.

Despite of these shortcomings, Regge pole motivated parametrizations are rather
successful in predicting high-energy cross sections
in the currently accessible energy range \cite{Donnachie92b,PDG00}. 
\begin{figure}[htb]
\centerline{
\epsfig{figure=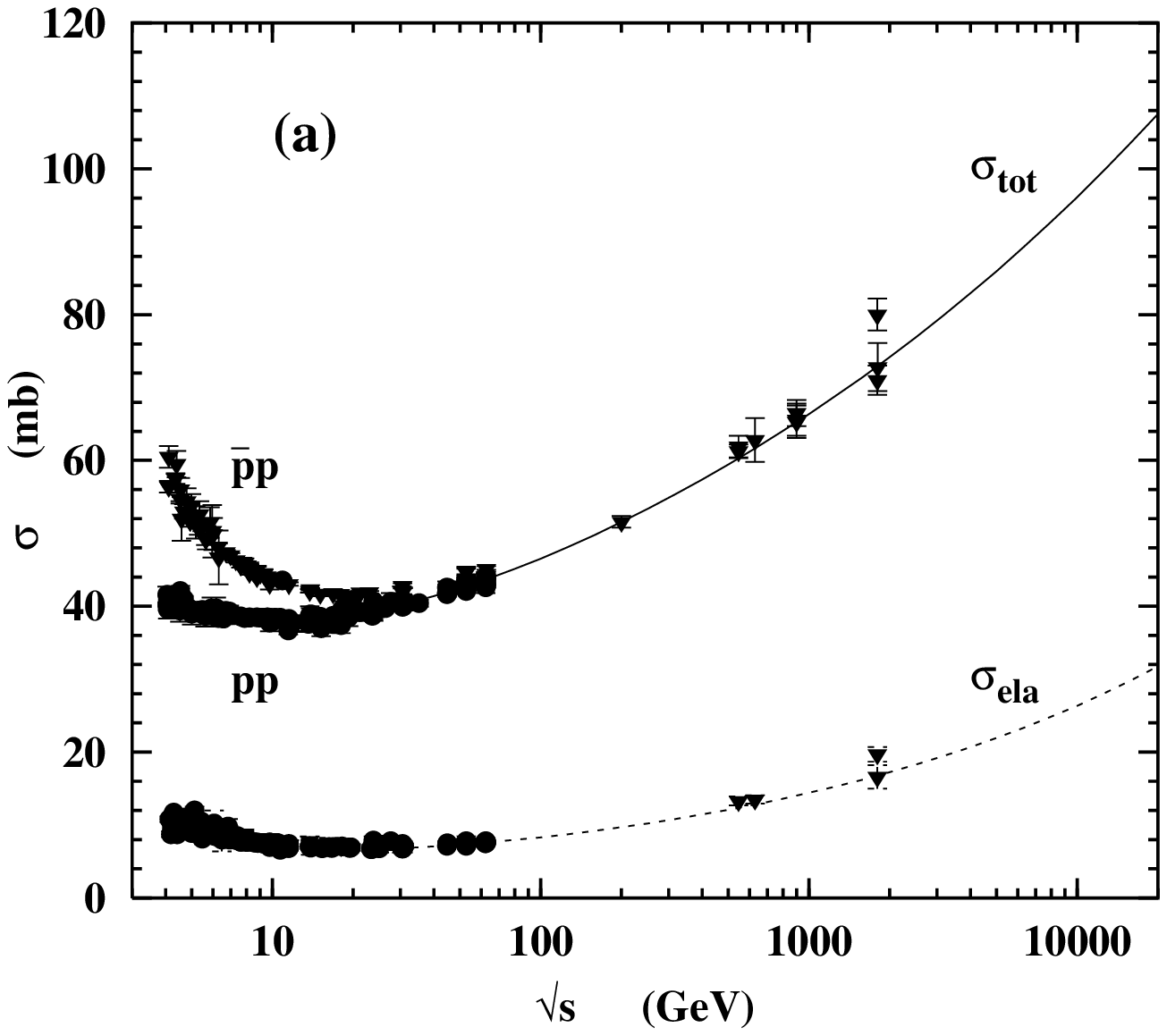,width=6cm}
\epsfig{figure=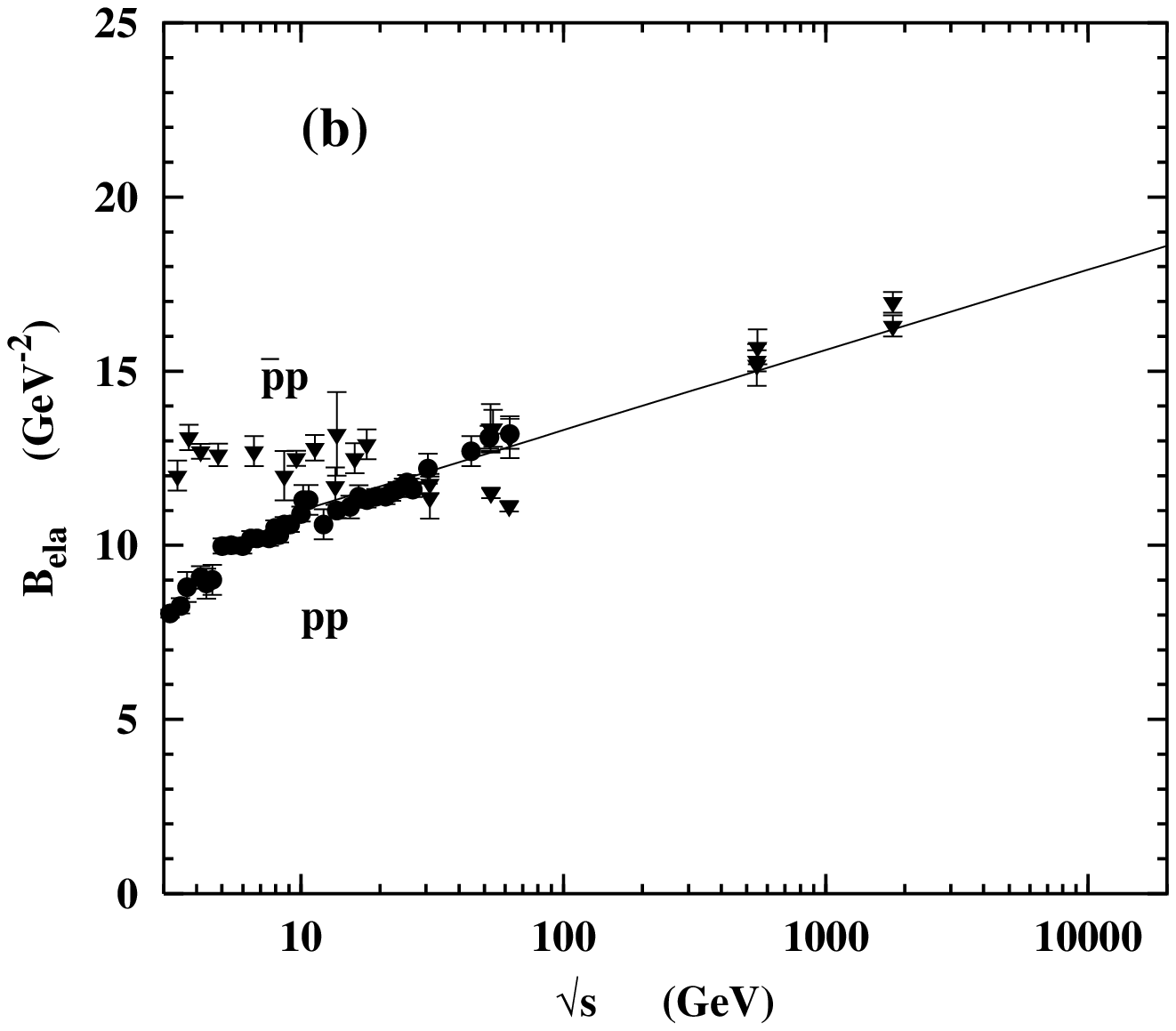,width=6cm}
}
\caption{Comparison of a simple Regge fit using two poles with $pp$ and
$p\bar p$ data. The parameters for the total cross section fit are 
from \protect\cite{Donnachie92b}.
\label{fig-pp-cs}
}
\end{figure}
For example, Fig.~\ref{fig-pp-cs}a shows the prediction by Donnachie and
Landshoff \cite{Donnachie92b} together with $pp$ and $p\bar p$ data. In
Fig.~\ref{fig-pp-cs}b we show the energy dependence of the elastic slope
parameter, $B_{\rm ela}$, which is also in agreement with the simple 
Regge pole expectation of a logarithmic energy dependence
(see Eq.~(\ref{cs-pole})).


\subsection{Multi-pomeron models}

As already mentioned, the pomeron exchange amplitude violates
unitarity. Although this doesn't seem to be a serious problem at current
collider energies, it indicates an inconsistency of this model.

The simplest way of overcoming this problem without abandoning 
Regge theory altogether is the introduction of multiple pomeron
exchanges in a single scattering process.
Assuming that large momentum transfers are suppressed by the dynamics 
of the strong interaction, as it is the case in QCD, 
multi-pomeron exchange can be described
on the basis of Gribov's Reggeon field theory 
(RFT) \cite{Gribov67a-e}.
Then the total amplitude can be written as the sum of $n$-pomeron exchange
amplitudes $A^{(n)}(s,t)$.
For each $n$-pomeron graph one can define a theoretical ``total'' cross
section applying the optical theorem to the corresponding $n$-pomeron
amplitude
\begin{equation}
\sigma^{(n)} = (-1)^{n+1} \frac{1}{s} \Im m \left( A^{(n)} \right),
\hspace*{1cm} \sigma_{\rm tot} = \sum_{n=1}^{\infty} (-1)^{n+1}
\sigma^{(n)}\ .
\end{equation}
Here the alternating sign has been introduced by definition to keep all
partial cross sections $\sigma^{(n)}$ positive.

\begin{figure}[!htb] \centering
\hspace*{0.25cm}
\psfig{file=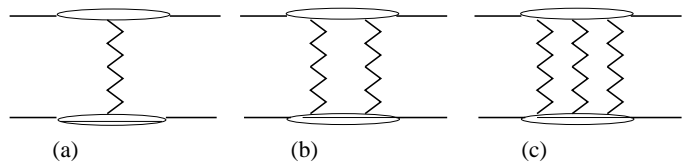,height=2cm}
\caption{
\label{allgra0} \em
Hadron-hadron scattering via pomeron exchange:
(a) one-pomeron,  (b) two-pomeron, and (c) three-pomeron exchange graphs.
\label{fig-multi-pom}
}
\end{figure}

As a simplified model we consider only the first two graphs shown in
Fig.~\ref{fig-multi-pom}, assuming
$\sigma^{(n)} \ll 4 \sigma^{(2)} < \sigma^{(1)}$ with $n > 2$.
Then, the total cross section reads 
$\sigma_{\rm tot} =  \sigma^{(1)} - \sigma^{(2)}$, where $\sigma^{(1)}$
and $\sigma^{(2)}$ are the cross sections of the one- and two-pomeron
exchange graphs, respectively.
The energy-dependence of the two-pomeron cross section is directly
linked to that of $\sigma^{(1)} \sim s^{\Delta}$ and turns out to be
$\sigma^{(2)} \sim s^{2\Delta}$.
The two-pomeron cross section grows faster with energy 
than the one-pomeron
cross section. This leads to a weaker
energy-dependence of the total cross section than in the single-pomeron
exchange model.

One particular realization of the multi-pomeron exchange is the eikonal
model using the pomeron amplitude as eikonal function. In general,
predictions of such models
are in good agreement with cross section data and are
nearly indistinguishable from the single-pomeron exchange model in the
energy range up to several TeV c.m.s. energy.


\subsection{Photon-induced hadronic interactions}

The generalization to photon-induced interactions is straight-forward
in terms of the generalized vector dominance model (GVDM)
\cite{Sakurai72b,Bauer78}. 
The photon is considered as a superposition of two
types of states, a bare photon and hadronic fluctuations.
With the probability of the order of the fine
structure constant $\alpha_{\rm em}$ the photon can be found in a
fluctuation which interacts like a hadron. Since the cross section of
the bare photon is small it can be neglected in most applications.
In the proton rest frame, the life time $t_f$ of
the hadronic fluctuation with momentum $k$ and mass $M$ is much larger
than the typical interaction time $t_i$
\begin{equation}
t_f \sim \frac{1}{\sqrt{M^2+k^2}-\sqrt{-Q^2+k^2}}
\approx \frac{2 k}{M^2+Q^2}\ ,
\label{eq:fluc-time}
\end{equation}
where $Q^2$ denotes the photon virtuality.
The hadronic amplitudes for the scattering of
longitudinally ($L$) and transversely ($T$) polarized
photons are written as
\begin{eqnarray}
A^{(T)}_{\gamma h \rightarrow X} &=&
\sum_{V}
\left(\frac{e}{f_V}\right)
\frac{m_V^2}{m_V^2-q^2-i\Gamma_V m_V}
\ A^{(T)}_{V h \rightarrow X}(s,t)
\label{eq:vdm-trans}
\\
A^{(L)}_{\gamma h \rightarrow X} &=&
\sum_{V}
\left(\frac{e}{f_V}\right)\left(\frac{-q^2
\xi_V}{m_V^2}\right)^{\frac{1}{2}}
\frac{m_V^2}{m_V^2-q^2-i\Gamma_V m_V}
\ A^{(T)}_{V h \rightarrow X}(s,t)\ ,
\label{eq:vdm-long}
\end{eqnarray}
with $q$, $m_V$, and $\Gamma_V$ being the four-momentum of the incoming
photon
and the
mass and the decay width of the vector meson $V$, respectively.
$A^{(T)}_{V h \rightarrow X}$ is the
amplitude for the scattering of the transversely polarized vector meson
$V$
off the hadron $h$. The parameter $\xi_V$ is introduced to allow for
different cross sections of transversely and  longitudinally polarized
vector mesons.

The GVDM is very successful in predicting the total and and quasi-elastic
vector meson photoproduction cross sections \cite{Abramowicz99a}.
However, the $s$-dependence of the $\gamma p$ and $\gamma\gamma$
cross sections changes considerably from real to virtual photons.
This cannot be understood on the basis of GVDM alone. Many alternative
models have been published.  The most promising calculations combine
within the color-dipole framework
ideas of GVDM with perturbative QCD calculations, obtaining a reasonable
description of the measurements
\cite{Badelek92}.


\subsection{Impact parameter picture of high-energy scattering}

The impact parameter picture of high-energy scattering is very useful
for discussing effects related to the wave character of the scattering
particles. Its meaning becomes most 
transparent by considering again the partial wave expansion of the
scattering amplitude
\begin{equation}
A(s,t) = 16\pi \sum_{l=0}^\infty
(2 l +1 )\;a_l(s)\;P_l(\cos \theta).
\label{partial-wave}
\end{equation}
In the classical (geometrical) description, the
$l$th term on the r.h.s. of Eq.~(\ref{partial-wave}) with angular
momentum $l$
is assumed to correspond to
the interaction of two particles with an impact parameter $\vec b$
satisfying $ l = |\vec b| k$.
The impact parameter vector $\vec b$ is by definition perpendicular
to the momentum $\vec k$ of the incoming particle.
At high energy the sum can be approximated by the integral
\begin{equation}
A(s,t) = 16 \pi \int_{0}^\infty d l\ (2 l +1 )
\;a_l(s)\;P_l(\cos \theta)\ ,
\label{partial1}
\end{equation}
using for $a_l(s)$ the analytic continuation in $l$.
Taking the limit 
$P_l(\cos\theta) \stackrel{l \rightarrow \infty}{\longrightarrow}
J_0\left[ (2l+1)\sin(\theta/2) \right]$ and substituting
\begin{equation}
J_0(z) = \frac{1}{2 \pi} \int_0^{2 \pi} d\varphi e^{iz \cos\varphi},
\hspace*{1cm}
\vec b\cdot\vec q = \vec b\cdot\vec{q}_\perp = b q_\perp\cos\varphi\ ,
\end{equation}
Eq.~(\ref{partial1}) becomes
\begin{equation}
A(s,t) = 4 s \int d^2 \vec b\ a(s,\vec b)
e^{i \vec{q}_\perp\cdot \vec b}\ ,
\hspace*{5mm} {\rm with} \hspace*{5mm}
a(s,\vec b) = a_l(s) \big|_{l = k B}\ .
\label{partial2}
\end{equation}
The impact parameter amplitude $a(s,\vec b)$ is
the Fourier transform of the elastic scattering amplitude.
Furthermore, in analogy to geometrical optics, the 
function $a(s,\vec b)$ can be
interpreted as the density function of sources of scattered waves
producing interference patterns.
The total and elastic cross sections are given by
\begin{equation}
\sigma_{\rm ela}(s) = 4 \int d^2\vec b\ | a(s,\vec b) |^2,
\hspace*{1cm}
\sigma_{\rm tot}(s) = 4 \int d^2\vec b\ \Im m(a(s,\vec b))\ .
\label{sigtot-imp}
\end{equation}

\begin{figure}[htb]
\centerline{
\begin{minipage}[t]{11.6cm}
  \begin{minipage}[t]{7cm}
    \epsfig{figure=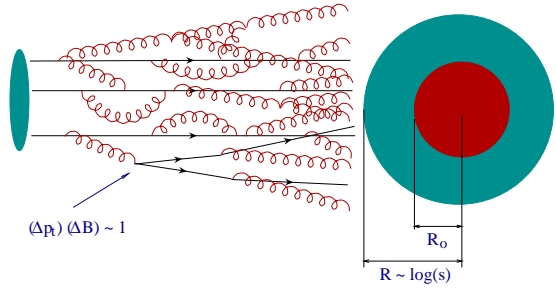,height=3.8cm}
  \end{minipage}
\hfill
  \begin{minipage}[b]{3.5cm}
    \caption{Partonic interpretation of the logarithmic growth of hadrons
    with energy.  
    \label{fig-low-x}
    }
  \end{minipage}
\end{minipage}
}
\end{figure}
For an amplitude with
exponential $t$-dependence $A(s,t) \sim i s \sigma_{\rm tot}
\exp\{\frac{1}{2} B_{\rm ela} t\}$ the corresponding impact parameter
amplitude reads
\begin{equation}
a(s,\vec b) = i \frac{\sigma_{\rm tot}}{8 \pi B_{\rm ela}}
\exp\left(-\frac{{\vec b}^2}{2 B_{\rm ela}}\right)\ .
\end{equation}
The width of the amplitude in $\vec b$-space (i.e.\ the ``gray'' area)
is given by the $t$-slope of the elastic
cross section and, according to Eq.~(\ref{cs-pole}), grows
logarithmically with $s$. This can also be understood in terms of 
partons in a hadron as shown in Fig.~\ref{fig-low-x}. 
Each parton splitting involves a
transverse momentum transfer which corresponds to a small displacement
in impact parameter. The total displacement of a parton follows from the
number of steps $N$ of its random walk, $R^2 \sim \Delta b\cdot N \sim
(\Delta p_\perp)^{-1} \ln s$. This picture also readily explains why the
elastic slope of hard processes has a much weaker energy dependence: the
displacement in $\vec b$ is much smaller because of the larger $\Delta
p_\perp$.

From the unitarity constraint
$|a(s,\vec b)| < 1$ one gets as high-energy limit $\sigma_{\rm tot} <
c \ln^2 s$ with $c$ being a constant. The special case of the black disk
limit (e.g. maximum absorption) corresponds to $a(s,\vec b) = i/2$ for
$b<R$ and $a(s,\vec b) = 0$ otherwise. Then
the inelastic cross
section (cross section for absorption) is given by the geometrical
size of the disk
$\sigma_{\rm ine} = \pi R^2$ whereas the total cross section is twice
the disk size, $\sigma_{\rm tot} = 2 \pi R^2$.
This is a result of unitarity: absorption gives rise to elastic
scattering $\sigma_{\rm ela} = \pi R^2 = \sigma_{\rm tot}/2$.


\section{Multiparticle production}


\subsection{Unitarity and cuts}

\begin{figure}[htb]
\centerline{
\begin{minipage}[t]{11.6cm}
  \begin{minipage}[t]{8cm}
    \epsfig{figure=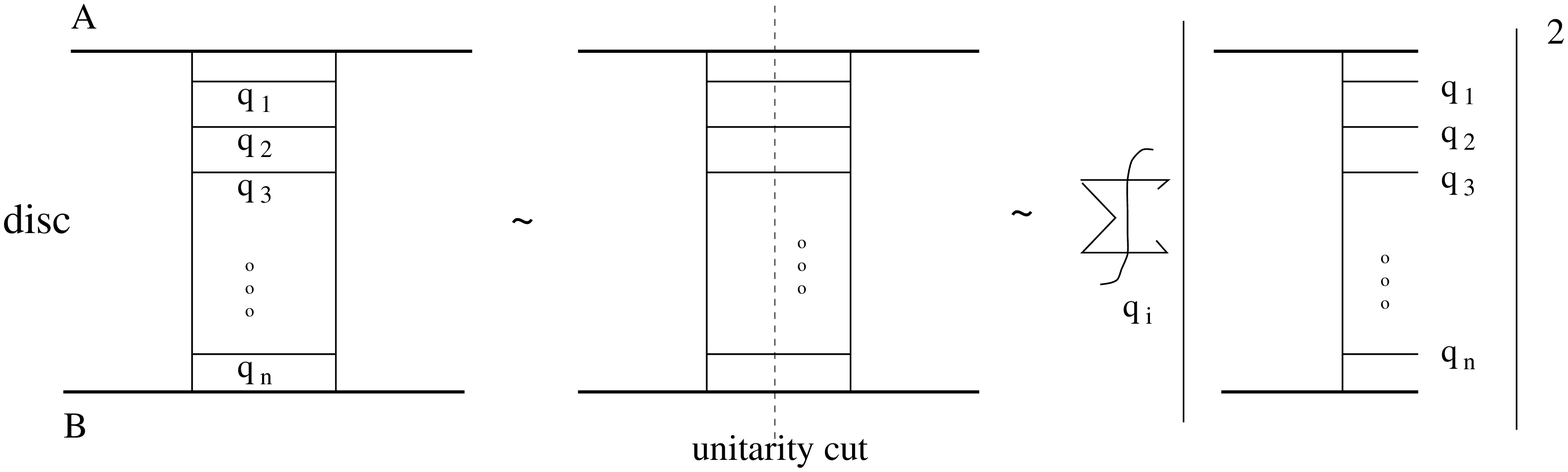,width=8cm}
  \end{minipage}
\hfill
  \begin{minipage}[b]{3.0cm}
    \caption{Graphical representation of optical theorem.
    \label{fig-unitarity}
    }
  \end{minipage}
\end{minipage}
}
\end{figure}

To link particle production to elastic scattering amplitudes one can
use the optical theorem. In terms of Feynman
diagrams a particle propagator collapses to an external on-shell 
particle line if its imaginary part is taken
\begin{equation}
\Im m\left(\frac{1}{p^2-m^2-i\epsilon}\right) = \pi \delta(p^2-m^2)\ .
\end{equation}
Therefore taking the discontinuity (i.e. the imaginary part) of a
propagator is commonly referred to as ``cutting''.
This is also the reason why, up to kinematical factors, 
the discontinuity of elastic
amplitude is equal to the squared matrix element describing particle
production, as sketched in Fig.~\ref{fig-unitarity}.

\begin{figure}[htb]
\centerline{
\epsfig{figure=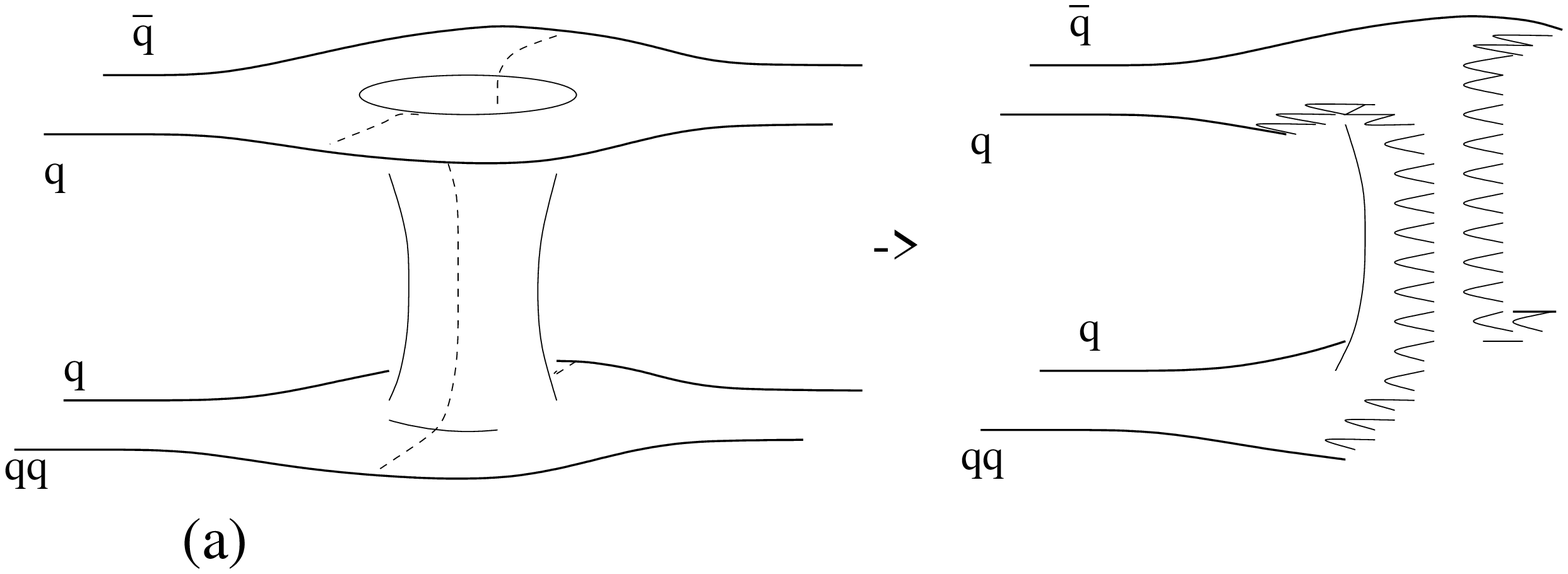,width=6.5cm}\hspace*{1cm}
\epsfig{figure=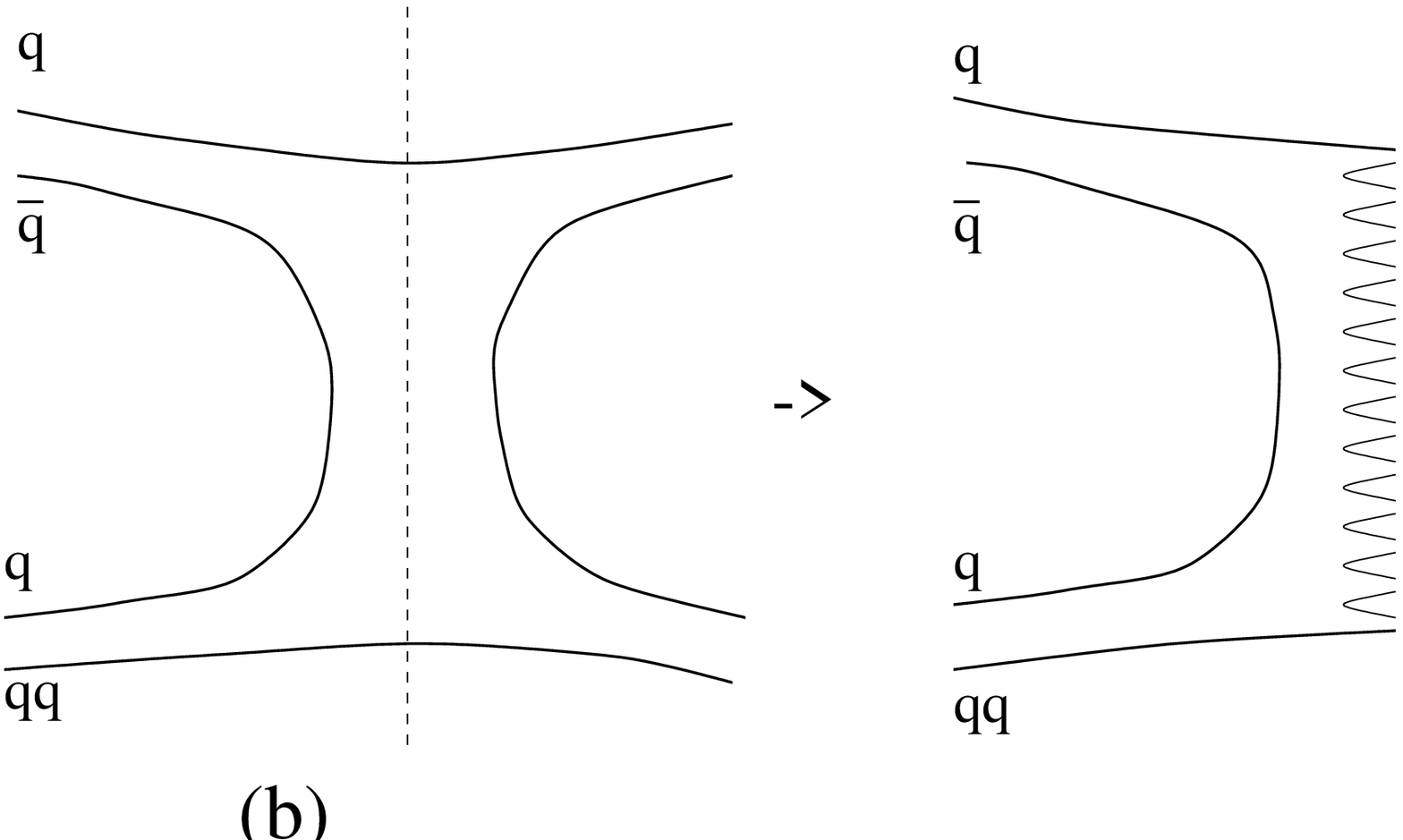,width=4.2cm}
}
\caption{Cuts and final state topologies of (a) pomeron and
(b) reggeon exchange graphs.
\label{fig-cut-pom}
}
\end{figure}

One particular application is the discussion of the topologies of
cut pomeron and reggeon amplitudes in the limit of large numbers of
colors and flavours in QCD \cite{Veneziano76}. As depicted in
Fig.~\ref{fig-cut-pom} the dominant final state contributions are a
two- and one-string configurations.


\subsection{Abramovski-Gribov-Kancheli cutting rules}

In general, there are many possibilities of cutting diagrams
involving multiple pomerons.
The dominant contribution of the different cut
configurations to the total discontinuity can be calculated with the
Abramovski-Gribov-Kancheli (AGK) cutting
rules \cite{Abramovski73-e}:\\
(i) The dominant contributions are given by
cut configurations where the cut involves all intermediate particle
states
of a pomeron (a contribution due to a partially cut pomeron is
sub-leading).
\\
(ii) For an $n$-pomeron
exchange graph, the contribution to the
discontinuity with $\nu$ cut pomeron propagators is
$B^n_\nu \ 2 \Im m\left( A^{(n)} \right)$. Here, $A^{(n)}$ denotes the
$n$-pomeron amplitude and $\nu$ is restricted to $0\le \nu \le n$.
The combinatorial factors (AGK weights) $B^n_{\nu}$ are
\begin{equation}
B_{\nu}^n = \left\{ \begin{array}{rcl}
(-1)^{\nu-1} \frac{\displaystyle n!}{\displaystyle \nu!(n-\nu)!}
2^{n-1} & : &\nu \ge 1\\
 & & \\
 1 - 2^{n-1}         & : & \nu = 0\ .
\end{array}
\right.
\label{AGK-factors}
\end{equation}
\\
(iii) The coefficients $B_{\nu}^n$ satisfy
$\sum_{\nu = 0}^{n}  B^n_\nu = 1$,
which means that all leading contributions to the total discontinuity
are included.

\begin{figure}[!htb]
\centering
\hspace*{0.25cm}
\psfig{file=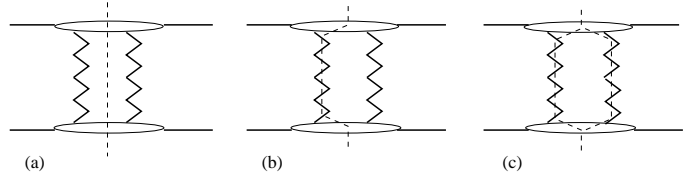,width=90mm}
\caption{
Breakdown of the total discontinuity of the two-pomeron exchange graph
according to the AGK cutting rules:
(a) the diffractive cut describing low-mass diffraction,
(b) the one-pomeron cut, and
(c) the two-pomeron cut.
\label{pom2-cut}
}
\end{figure}

For example, there are three leading cuts of the two-pomeron graph
(Fig.~\ref{fig-multi-pom}b):
the diffractive cut with the weight -1
(Fig.~\ref{pom2-cut}a), the one-pomeron cut with the weight 4
(Fig.~\ref{pom2-cut}b),
and the two-pomeron cut with the weight -2 (Fig.~\ref{pom2-cut}c).
The cross sections for the different final states of the two-pomeron
exchange graph are (a) diffractive cut: $\sigma_{\rm diff} = \sigma^{(2)}$,
(b) one-pomeron cut: $\sigma_{1} = -4 \sigma^{(2)}$, and
(c) two-pomeron cut: $\sigma_{2} = 2 \sigma^{(2)}$, see
Fig.~\ref{pom2-ine}.
Note that the diffractive cut of the two-pomeron graph gets a negative
AGK weight, hence giving in total a positive, experimentally observable
cross section.
However, the cross section for the
one-pomeron cut of the two-pomeron graph is negative.
Since the one-pomeron cut of the
one-pomeron graph has the same inelastic final state as the
one-pomeron cut of the two-pomeron graph (Fig.~\ref{pom2-ine}b), 
one can sum both
contributions to obtain a positive cross section.

Let's consider again the example of the one- and and two-pomeron graphs.
Concerning the topologies of the final state particles,
the total cross section is built up of
the sum of the partial cross sections of the one- and two-pomeron
exchange graphs:
\begin{figure}[!htb]
\centering
\hspace*{0.25cm}
\psfig{file=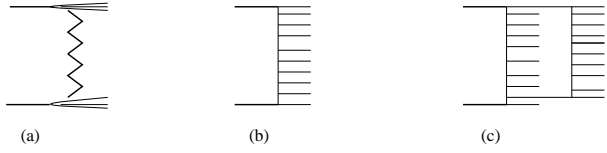,width=80mm}
\caption{
Inelastic final states resulting from 
(a) the diffractive cut describing low-mass diffraction,
(b) the one-pomeron cut, and
(c) the two-pomeron cut.
\label{pom2-ine} 
}
\end{figure}
(i) one-pomeron cut 
$\sigma_1 =\sigma^{(1)}-4\ \sigma^{(2)}$,
(ii) two-pomeron cut 
$\sigma_2 =2\ \sigma^{(2)}$, and
(iii) diffractive cut of the two-pomeron graph 
$\sigma_{\rm diff} = \sigma^{(2)}$.

The particle density in pseudorapidity due to a one-pomeron cut
is assumed to be almost energy independent (which is true for
longitudinal phase space models) and is denoted by $dN_1/d\eta$.
Provided the two-pomeron cut gives two times the particle
yield as compared to the one-pomeron cut (in central pseudorapidity
region, see Fig.~\ref{pom2-ine}),
the inclusive inelastic charged particle cross section is given by
\begin{equation}
\frac{d\sigma_{\rm ch}}{d\eta}\bigg|_{\eta\approx 0}
= 1\times \sigma_1 \frac{dN_1}{d\eta}
+ 2\times \sigma_2 \frac{dN_1}{d\eta}
+ 0\times \sigma_{\rm diff} \frac{dN_1}{d\eta}
= \sigma^{(1)}\ \frac{dN_1}{d\eta}\ .
\label{AGK-cancellation}
\end{equation}
Due to the topology of diffractive final states, almost
no particles are produced in the central region in the case of a
diffractive cut.
Note that only the one-pomeron graph
determines the inclusive particle cross section in the central region
(AGK cancellation).

On the other hand, the inclusive charged particle density follows from
\begin{equation}
\frac{dn_{\rm ch}}{d\eta}\bigg|_{\eta\approx 0} =
\frac{\sigma^{(1)}}{\sigma_{\rm tot}}\
\frac{dN_1}{d\eta}\bigg|_{\eta\approx 0}
\approx
\frac{\sigma^{(1)}}{\sigma^{(1)} - \sigma^{(2)}}\
\frac{dN_1}{d\eta}\bigg|_{\eta\approx 0}\ .
\label{rho0-tot}
\end{equation}
Eq.~(\ref{rho0-tot})
allows us to understand the observed behaviour of $\rho(0)=dn_{\rm ch}/d\eta$
in $pp$ and $p\bar p$ collisions.
With $\sigma^{(1)} \sim s^\Delta$ and $\sigma_{\rm tot} \sim s^{0.08}$,
one gets a power-law increase of the central particle density.
This is confirmed by experiment\cite{Geich-Gimbel89a}: 
$\rho(0) \approx 0.74\ s^{0.105}$.

Since the two-pomeron graph has a cross section which
increases faster with energy than the cross section of one-pomeron
graph, the model predicts also an increase
of the multiplicity fluctuations with increasing collision energy.
In a model without geometric scaling this leads to violation of KNO
scaling\cite{Koba72a} at high energies.
Furthermore, due to the characteristic structure of the one- and 
two-pomeron cuts,
strong long-range correlations in pseudorapidity are naturally
explained. For a detailed discussion, see for example \cite{Capella94a}
and Refs.\ therein.
\begin{figure}[htb]
\centerline{
\unitlength1mm
\begin{picture}(120,55)
\put(0,0){\psfig{file=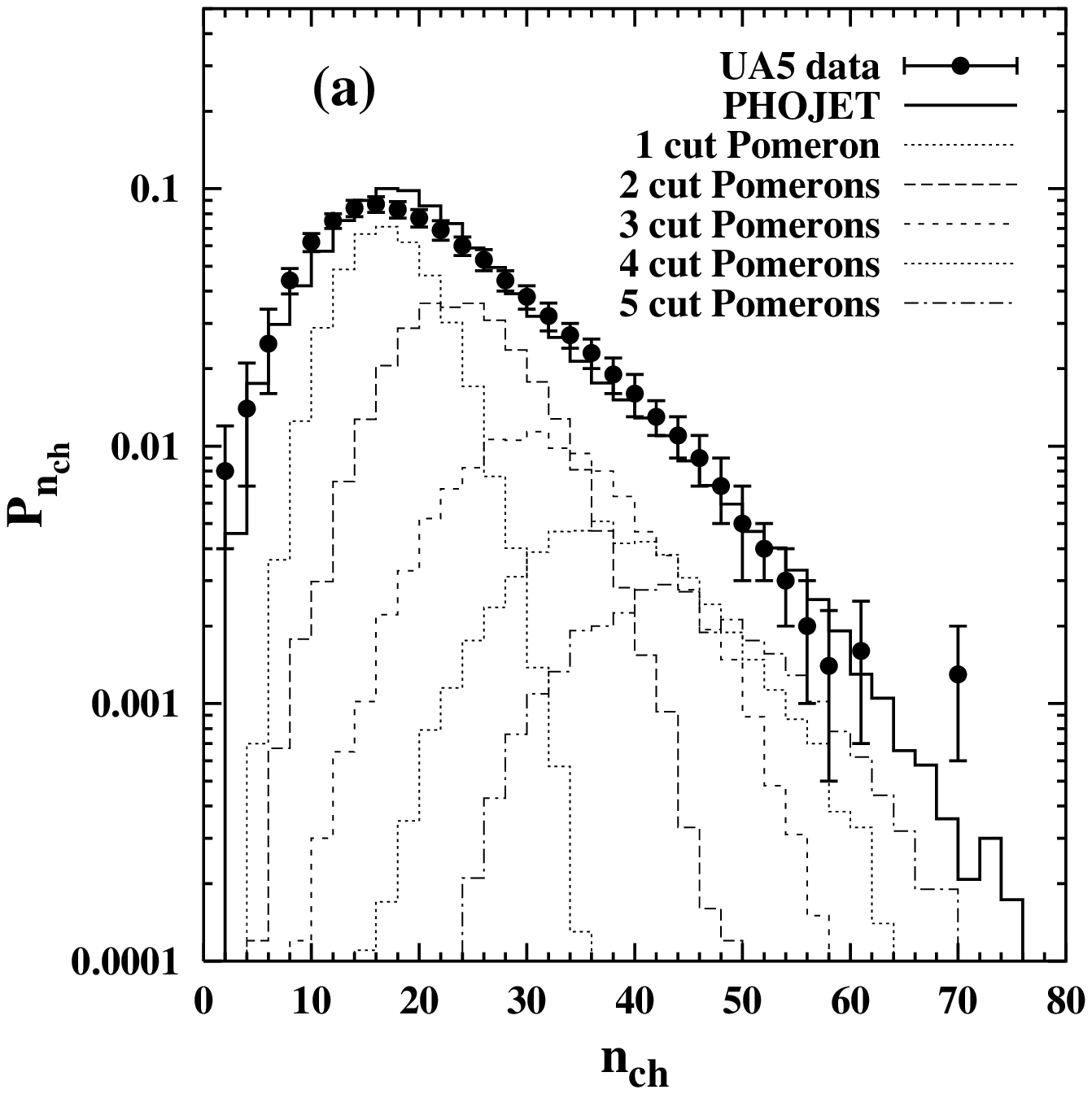,width=5.9cm}}
\put(60,0){\psfig{file=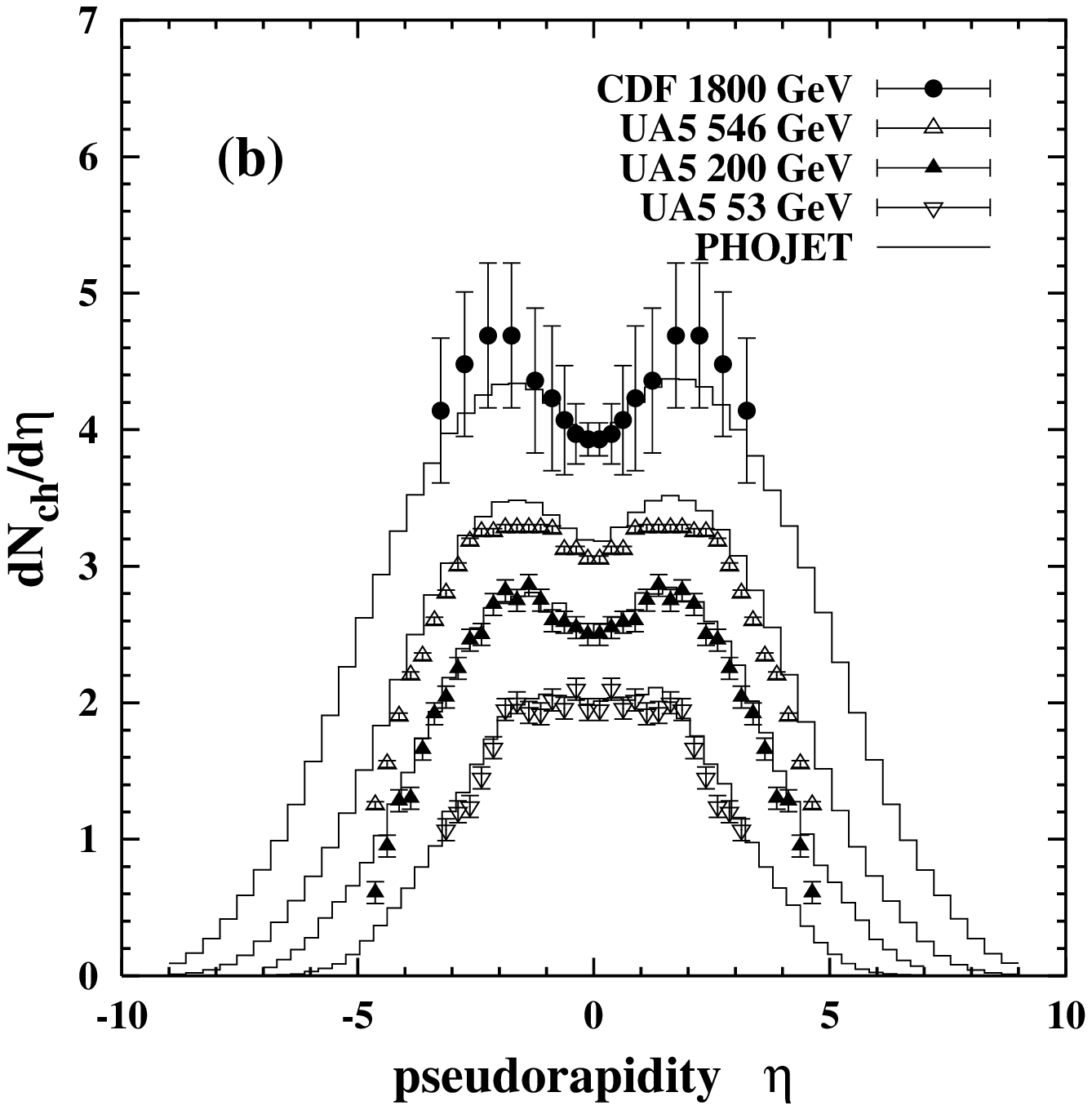,width=5.9cm}}
\end{picture}
}
\caption{
(a) Decomposition of the multiplicity distribution in $p\protect\bar{p}$
collisions according to the number of generated pomeron cuts
at $\protect\sqrt s$ = 200 GeV. 
(b) Energy-dependence of charged particle pseudorapidity density in $p\bar p$
collisions.\label{mulpom}
}
\end{figure}

Finally, as a representative example, model predictions for the
multiplicity distribution and the pseudorapidity distribution 
(calculated with the
{\sc Phojet} event generator\cite{Engel96j}) are compared with
collider data\cite{Geich-Gimbel89a} in Fig.~\ref{mulpom}.


\section{Summary}

Regge theory, AGK cutting rules, QCD in the limit of large numbers of
colors and flavours, and the geometrical
interpretation of high-energy scattering prove very useful in
understanding the basic properties of cross sections and multiparticle
production. They are some of the tools available today to study soft
processes theoretically. Many models combine them to obtain a detailed 
description of soft processes
\cite{Capella94a,Kaidalov86c,Aurenche92,Engel96j,Drescher01a}.

However we are far from being able to reliably
calculate predictions for most soft production processes. All approaches
or models discussed here have severe shortcomings. For example,
Regge theory does not 
predict transverse momentum-related quantities, and most models
implement AGK cutting rules without energy-momentum conservation,
to name but a few.
Furthermore the rather large number of free parameters limits the 
predictive power of calculations.

\vspace*{2mm}
\noindent
{\bf Acknowledgements:}
This work is supported by U.S. Department
of Energy grant DE-FG02~91ER~40626. The author is
grateful to L.\ Frankfurt, T.K.\ Gaisser, J.\ Ranft, S.\ Roesler, T.\
Stanev, and M.\ Strikman for many interesting discussions.



\begin{thebibliography}{10}

\bibitem{Sachrajda01a}
C.~T. Sachrajda:
\newblock Phenomenology from lattice QCD,
\newblock (hep-ph/0110304),
\newblock 2001

\bibitem{Khoze01a}
V.~A. Khoze and W.~Ochs:
\newblock Theory of multiparticle production in the soft region,
\newblock MPI-PHT-2001-42 (hep-ph/0110295),
\newblock 2001

\bibitem{Kaidalov79}
A.~B. Kaidalov:
\newblock Phys. Rep. 50 (1979) 157

\bibitem{Collins82a}
P.~D.~B. Collins and A.~D. Martin:
\newblock Rep. Prog. Phys. 45 (1982) 335

\bibitem{Capella94a}
A.~Capella, U.~Sukhatme, C.~I. Tan  and J.~Tr\^an Thanh~V\^an:
\newblock Phys. Rep. 236 (1994) 225

\bibitem{Abramowicz97b}
H.~Abramowicz, L.~Frankfurt  and M.~Strikman:
\newblock Surveys High Energ. Phys. 11 (1997) 51

\bibitem{Levin98b}
E.~Levin:
\newblock An introduction to pomerons,
\newblock (hep-ph/9808486),
\newblock 1998

\bibitem{Kaidalov99}
A.~B. Kaidalov:
\newblock Surveys High Energ. Phys. 13 (1999) 265

\bibitem{Collins77}
P.~D.~B. Collins:
\newblock {\em An Introduction to Regge Theorie \& High Energy Physics,}
\newblock Cambridge University Press, Cambridge 1977

\bibitem{Donnachie92b}
A.~Donnachie and P.~V. Landshoff:
\newblock Phys. Lett. B296 (1992) 227

\bibitem{PDG00}
Particle Data Group:  D.~E. Groom et~al.:
\newblock Eur. Phys. J. C15 (2000) 1

\bibitem{Gribov67a-e}
V.~N. Gribov:
\newblock Sov. Phys. JETP 26 (1968) 414;
M.~Baker and K.~A. Ter-Martirosyan:
\newblock Phys. Rep. 28C (1976) 1

\bibitem{Sakurai72b}
J.~J. Sakurai and D.~Schildknecht:
\newblock Phys. Lett. 41B (1972) 489

\bibitem{Bauer78}
T.~H. Bauer, R.~D. Spital, D.~R. Yennie and F.~M. Pipkin:
\newblock Rev. Mod. Phys. 50 (1978) 261; erratum Rev. Mod. Phys. 51
(1979) 407

\bibitem{Abramowicz99a}
H.~Abramowicz and A.~Caldwell:
\newblock Rev. Mod. Phys. 71 (1999) 1275

\bibitem{Badelek92}
B.~Badelek and J.~Kwieci\'nski:
\newblock Phys. Lett. B295 (1992) 263;
A.~D. Martin, M.~G. Ryskin  and A.~M. Stasto:
\newblock Eur. Phys. J. C7 (1999) 643;
E.~Gotsman, E.~Levin, U.~Maor  and E.~Naftali:
\newblock Eur. Phys. J. C10 (1999) 689;
M.~McDermott, L.~Frankfurt, V.~Guzey  and M.~Strikman:
\newblock Eur. Phys. J. C16 (2000) 641

\bibitem{Veneziano76}
G.~Veneziano:
\newblock Nucl. Phys. B117 (1976) 519

\bibitem{Abramovski73-e}
V.~A. Abramovski, V.~N. Gribov  and O.~V. Kancheli:
\newblock Sov. J. Nucl. Phys. 18 (1974) 308

\bibitem{Geich-Gimbel89a}
C.~Geich-Gimbel:
\newblock Int. J. Mod. Phys. A4 (1989) 1527

\bibitem{Koba72a}
Z.~Koba, H.~B. Nielsen  and P.~Olesen:
\newblock Nucl. Phys. B40 (1972) 317

\bibitem{Engel96j}
R.~Engel and J.~Ranft:
\newblock Phys. Rev. D54 (1996) 4244

\bibitem{Kaidalov86c}
A.~B. Kaidalov and O.~I. Piskunova:
\newblock Z. Phys. C30 (1986) 145

\bibitem{Aurenche92}
P. Aurenche {\it et al.}:
Phys. Rev. D45 (1992) 92; S. Roesler, R. Engel and J. Ranft:
The Monte Carlo event generator DPMJET-III (hep-ph/0012252)

\bibitem{Drescher01a}
H.~J. Drescher, M.~Hladik, S.~Ostapchenko, T.~Pierog  and K.~Werner:
\newblock Phys. Rep. 350 (2001) 93

\end{thebibliography}


\end{document}